\documentclass{sig-alternate}

\usepackage{url}
\usepackage{enumitem}
\usepackage{comment}

\begin{document}

%

\title{Code Drones}

\numberofauthors{1}
\author{
\alignauthor
Mithun P. Acharya$^1$, Chris Parnin$^2$, Nicholas A. Kraft$^1$, Aldo Dagnino$^1$, Xiao Qu$^1$\\
       \affaddr{$^1$ABB Corporate Research, Raleigh, North Carolina, USA}\\
       \affaddr{$^2$North Carolina State University, Raleigh, North Carolina, USA}\\
       \email{\{mithun.acharya,nicholas.a.kraft,aldo.dagnino,xiao.qu\}@us.abb.com, chris.parnin@ncsu.edu}
}

\maketitle
\begin{abstract}
We propose and explore a new paradigm called \texttt{Code Drones} in which every software artifact such as a class is an intelligent and socially active entity. In this paradigm, \textit{humanized} artifacts take the lead and choreograph (socially, in collaboration with other intelligent software artifacts and humans) automated software engineering solutions to a myriad of development and maintenance challenges,  including API migration, reuse, documentation, testing, patching, and refactoring. We discuss the implications of having social and intelligent/cognitive software artifacts that guide their own self-improvement.
\end{abstract}




\section{Introduction}
\label{sec:intro}

Millions of classes\footnote{A software artifact is any tangible by-product of software development such
as a requirements document, UML diagram, class, package, test case, or a bug report. The ideas discussed in this paper may apply to any software artifact, but we discuss our ideas using a class as an exemplar software artifact.} sit idly for years in the version control systems doing nothing but change according to the whims and fancies of developers\textemdash they are treated as mere objects (pun intended). Well, no more. Enter a new world of \texttt{Code Drones}\footnote{We overload the term ``Code Drones''. \texttt{Code Drones} (when used with \texttt{typewriter} font) means the new paradigm we introduce in this paper. Code Drones (when used with normal font) means the artifacts that are social and intelligent by virtue of our paradigm. Thus, we can have a variety of code drones such as requirement drones, class drones, library drones, test drones, and bug drones.} where classes, our new human-like friends, are highly social creatures. Classes form Facebook-like social networks with their fellow classes and humans, tweet, email, instant message, and even apply for IPv6 addresses and make phone calls. These classes take great pride in their design and correctness and even have LinkedIn-like profiles to boast about their achievements. The classes, apart from being social, are highly intelligent as well. For example, classes learn, reason, and exercise their freedom of tweet/speech and democratic voting rights in deciding how they, or the APIs they use, evolve.  Like humans, these classes continuously browse social networks and the web to proactively find opportunities for self-improvement, both as individuals and as socially responsible community members. In doing so classes choreograph efficient solutions to many software development and maintenance challenges such as API migration, reuse, documentation, testing, patching, and refactoring. Later, driven by their human-like tendencies, classes frequently publish their improvements and accomplishments on their LinkedIn profiles, webpages, and social networks. Millions of these social and intelligent classes, i.e., code drones, will join forces with developers to create better and more reliable software\footnote{A shorter version of this draft to appear at ICSE Visions 2025 and Beyond, 2016}. 

The \texttt{Code Drones} paradigm is most closely related to the Multi-Agent Systems (MAS) paradigm~\cite{mas}. \textit{Agents} are sophisticated and intelligent computer programs that act autonomously on behalf of their users or another program across open and distributed environments~\cite{cmu_mas}. MAS applications cover a variety of domains including e-commerce, military logistics planning, supply-chain management, and financial portfolio management. Companies such as Netflix currently use MAS to assist in reliability of production systems. Multiple agents in the \emph{Simian Army}~\cite{chaos-monkey}, such as \emph{Chaos Monkey} and \emph{Doctor Monkey}, mutate production environments (e.g., shutting down services, introducing latency, or patching unhealthy instances) to perform resilience testing.

In the \texttt{Code Drones} paradigm, we apply MAS concepts to the domain of software engineering. Unlike the traditional MAS used by Netflix and others, where an agent represents a user or another program, in \texttt{Code Drones}, an agent represents a software artifact. In \texttt{Code Drones}, every software artifact has a dedicated agent whose sole goal is to improve the artifact it represents and thus the software of which the artifact is a part of. Agents achieve this goal by collaborating with other agents as well as humans. For example, with \texttt{Code Drones}, a class, through its agent, constantly gathers facts, \emph{reasons}, and \emph{auto-evolves} its code based on the \emph{experience} it gains by searching news about security vulnerabilities, reading tweets about new performant APIs, and correlating the test environment results with production Key Performance Indicators (KPIs). Can we build (semi)\textit{Autonomic Systems}~\cite{autocomp} for software development and maintenance with \texttt{Code Drones}?
 
As complexity increases and technical debt mounts, developers will have continued problems in maintaining, deploying, and stabilizing software.
We believe \texttt{Code Drones} can support developers in automating several software engineering tasks. Consider two example scenarios enabled by \texttt{Code Drones}:
\begin{description}[leftmargin=0em,topsep=0.25em,partopsep=0em,parsep=0.25em,itemsep=0em]
\item[Automated Pull Requests]{Tweets/telemetry raise speed concerns, triggering an automated analysis. A class ``realizes'' it is responsible and then submits a fix in a pull request. A developer decides whether or not to accept it.}
\item[Social Libraries]{A developer cannot choose from the 100s of JavaScript APIs available for a task. With \texttt{Code Drones} the developer finds an API whose ``LinkedIn profile'' shows connections to many popular projects. Its ``Facebook page'' lists an achievement of zero exceptions thrown in the last 90 days.}
\end{description}

The rest of the paper is organized as follows. In Section~\ref{sec:arch}, we outline the architecture for enabling the \texttt{Code Drones} paradigm. In Section~\ref{sec:apps}, we list the various applications of the \texttt{Code Drones} paradigm to software engineering problems. Finally, in Section~\ref{sec:discuss}, we discuss the various research challenges and opportunities moving forward with the \texttt{Code Drones} paradigm.

\section{Architecture}
\label{sec:arch}
In this section, we provide some initial directions for architecting and enabling the \texttt{Code Drones} paradigm. Enabling \texttt{Code Drones} will not require any changes to the classes themselves. Instead, each class is assigned a dedicated agent implemented using the \texttt{Code Drones} APIs (\textit{traditional}, \textit{social}, \textit{search}, and \textit{reasoning}, which we discuss in this section). These agents collaborate with other artifact agents and humans in evolving their owner classes and automating various software engineering tasks.  

The agents are implemented using the \texttt{Code Drones} APIs on a private or public cloud called the \texttt{Code Drones} agent cloud. An agent may modify its owner class and has access to its owner class' repositories, testing, and production environments. Further, an agent has full access to the Internet and any private or company intranets on which its owner class resides. Agents communicate with the developers using \textit{immersive} IDEs (See Section~\ref{sec:apps}) and human-artifact social networks. As the implementation of agents and agent cloud is hidden from the humans, to a developer it appears that he is directly communicating with the \textit{humanized} class rather than its agent. \textbf{Thus, in the rest of the paper, we will use the terms class and its dedicated \texttt{Code Drones} agent interchangeably.}

The \texttt{Code Drones} agents can be implemented at different granularities. The granularity ranges from having just one top-level intelligent agent at the software, library, or framework granularity to having thousands of intelligent \textit{microagents} -- one for each artifact -- at the artifact granularity. We argue that the \textit{Separation of Concerns} design principle applies to artifact intelligence as well, favoring agent implementation at the artifact granularity. With artifact granularity, agents are very specific to the artifacts they represent, thus enabling what we call as \textit{microintelligence} or \textit{microcognition}. However, the choice of granularity for the agent implementation will depend on the software. For some software packages, we may have a dedicated agent for each class in the package along with a master agent at the package granularity, coordinating the class agents. For some libraries, a single master agent may suffice in lieu of individual agents for each API or class in the library. 

In architecting \texttt{Code Drones}, we describe a working set of principles that guides our initial exploration:
\begin{itemize}[itemsep=0.2em,parsep=0em,topsep=0.2em]
\item An entity may change itself only when it has data to support its decision.
\item An entity shall not interact with others it should not, and shall not create overly excessive number of relationships (recommend using \textit{Dunbar's number}).
\item An entity shall perform actions at an observable cadence, such that it does not do actions that exceed an overseer's ability to regulate its behavior.
\item An entity shall be willing to cease existence.
\item An entity's intelligence shall not be restricted to passive nodes: It is not a participant in a neural network, it is closer to a company of workers.
\end{itemize}

Next, we discuss the \texttt{Code Drones} APIs, which every agent will have access to. Though the APIs are presented and discussed separately, classes (agents) use these APIs in tandem as a set of \textit{microservices}. 

\subsection{Traditional}
\label{subsec:traditional}

The traditional APIs enable the classes to use traditional software engineering approaches such as change impact analysis, symbolic execution, fault localization, and static verification. With access to the test and production environments, a class may infer and auto-repair a faulty member method using fault localization APIs that consider passing and failing test cases~\cite{jones:ase05}. Using symbolic or concolic execution APIs on the agent cloud, a class can test itself for security vulnerabilities \cite{bounimova:icse13}. Using program verification APIs, classes may collaborate with each other to check that the software they are all a part of satisfies critical behavioral properties thus ensuring reliability and correctness \cite{ball:cav07}. Many traditional software engineering approaches such as the aforementioned are not known to scale well and hence, companies today are implementing these as cloud services \cite{bounimova:icse13}. \texttt{Code Drones} follows the lead by implementing these traditional APIs on the agent cloud. 

The traditional APIs automate the many tasks that humans manually perform today when applying existing software engineering methodologies on software artifacts. Humanized artifacts behave like living organisms in many ways, and hence, software engineering paradigms inspired from the world of living organisms such as genetic programming~\cite{becker:13} are a natural fit for \texttt{Code Drones}. \texttt{Code Drones} implements genetic algorithm, mutation testing, and other paradigms inspired from living organisms such as automatic repair~\cite{kim:icse13}, self modification, and clone detection as traditional APIs that run on the agent cloud, readily available to the classes. 



\subsection{Social}
\label{subsec:social}

The social APIs enable the classes to form social networks with other artifacts and humans. Repositories such as GitHub, Q\&A sites such as Stack Overflow, and Codebook \cite{codebook} can be viewed as social networks of both people and software artifacts. But in GitHub and Stack Overflow, software artifacts 
are at best second-class citizens. In Codebook, unlike how people behave in classic social networks such as Facebook, software artifacts are \textit{passive} and \textit{non-intelligent} entities that do not actively participate in the network.   
By promoting classes to the first-class status and enabling the classes to proactively and intelligently participate, these social networks can be extended to serve as an early prototype that implements the social platform for the \texttt{Code Drones} paradigm.

With the social APIs, classes add their parent, sibling, and children classes into their family network. Packages, libraries, team, and products will all have fan pages which the classes and developers may like. The evolution history of a class will be published to its timeline. Classes write recommendations on the LinkedIn profile of their peer classes they have come to trust over time. Classes may also unfriend inactive or bug-prone classes. Classes also frequently search and browse (see search APIs, discussed next) the LinkedIn and Facebook pages of its APIs and developers on the social network. A class may choose not to allow developers with bad reputation to change them.
 
\subsection{Search}
\label{subsec:search}

The search APIs enable the classes to search source code repositories, the Web, and the social network. For example, making a Boa-like infrastructure \cite{boa} available to each class enables all classes to search ultra-large scale repositories and ask a range of questions. With Google/Stack Overflow-like search services, classes will harness the power of the Internet and their social network of developers and other classes. 

Developer Assistant \cite{devassist} is a Visual Studio addin from Microsoft that enables developers to search and reuse millions of code snippets and sample projects from within the Visual Studio IDE. The Developer Assistant uses Microsoft Bing \textit{contextual} search APIs to assist developers with compiler errors, API usage, and ''\textit{How do I}'' type questions/search. The \texttt{Code Drones} search APIs are similar to Developer Assistant's search APIs with two key differences. First, it is the classes (through their agents) that proactively invoke the search APIs and not the developers. Second, each class will have a \textit{personal} search assistant realized via its dedicated agent. Next, we discuss how a class may generate search queries and interpret the search results. 

A class may search the Internet with an API it uses as the keyword. If the search results indicate that the API is vulnerable, the class will immediately drop the API and either use an alternate API or an updated and a secure version of the vulnerable API. Existing approaches take a class and generate a natural language summary for it \cite{mcburney:icpc14}. The search queries can then be constructed or inferred from these summaries. Search and reasoning APIs (discussed next) are often used together for complex searches, as the classes will neither know what to search for nor be able to interpret the search results.
\subsection{Reasoning}
\label{subsec:reasoning}

The reasoning APIs are the most complex of all \texttt{Code Drones} APIs. Reasoning APIs give the classes the power of (artificial) intelligence. Classes will have access to AI services such as learning, reasoning, speech recognition/synthesis, and Watson-like \cite{watson} natural language analysis through the agent cloud. Classes can then be seen as powerful autonomous agents with self-improvement goals. Each class will analyze itself, ask questions on its social network, search the Internet, and then proactively use the gathered information for self-improvement. With the reasoning APIs, classes gain \textit{experience} and evolve over time. These humanized classes make optimal, and at times selfish, choices.

The reasoning APIs are complex to implement as they have to draw inferences based on software engineering data (via traditional APIs), social structure (via social APIs), and search results from various sources (via search APIs). Single and multi-agent learning is a well studied topic in MAS and will be central to the implementation of reasoning APIs. Other MAS concepts such as \textit{intentions}, \textit{know-how}, \textit{knowledge}, \textit{memory}, \textit{strategy}, \textit{beliefs}, and \textit{voting} will also play a key role in designing the reasoning APIs. For starters, however, even simple IFTTT (If This Then That) APIs without any reasoning power will let code drones tackle several software engineering problems efficiently, when used with other \texttt{Code Drones} APIs.

Implementing the aforementioned APIs and enabling the \texttt{Code Drones} paradigm will require concerted and sustained engineering efforts integrating research results from diverse areas of computer science. These areas include artificial intelligence, game theory, social networks analysis, cloud computing, machine learning, natural language analysis, and software engineering. However, given the recent advances in search, artificial intelligence, and cloud technologies, the era of humanized classes, i.e., code drones, harnessing the power of cloud and regularly collaborating with humans for composing all software is not far away. With today's technology, it should be straightforward to implement scenarios such as \textit{Automated Pull Requests} and \textit{Social Libraries} (discussed in Section~\ref{sec:intro}).

\section{Applications}
\label{sec:apps}

With \texttt{Code Drones}, all software artifacts will be proactive and aware, simplifying several software engineering problems from requirements to maintenance. We have begun to explore implementations of API drones~\footnote{\url{https://github.com/alt-code/ApiMonkey}} and test drones~\footnote{\url{https://github.com/alt-code/CrashMonkey}}. In addition to the two scenarios described in Section~\ref{sec:intro}, we envision many other applications of the \texttt{Code Drones} paradigm to software engineering problems.

\begin{description}[leftmargin=0em,topsep=0.25em,partopsep=0em,parsep=0.25em,itemsep=0em]
\item[API Migration]{Classes often depend on libraries, which may break at any time. For example, an application deployed using \emph{libusb1.0.8} may break if a developer is trying to compile with \emph{libusb1.0.9}. Instead of passively waiting for issue reports to surface such problems, a library drone continuously monitors new versions of the library that a class is dependent on and attempts to build and test the application to check for breaking changes. If possible, it patches itself to handle simple upgrade changes such as function renamings. A more sophisticated library drone may ask its friends if it has been able to successfully migrate between API versions in order to learn migration strategies.}

\item[Mining Software Repositories]{With human-artifact social networks, the complex relationships and dependencies among developers, software artifacts, and organization structure (that are often deeply buried or lost in source repositories) are evident and browsable by classes and humans alike. The large corpus of class and developer tweets and the network itself lends nicely to social network analysis~\cite{sna} and advanced network and tweet visualizations with graph analytics. \textit{Expertise graphs} and \textit{reputation systems} for developers and classes can be constructed with such network/graph analysis and visualizations, which may further be used for tasks such as automated bug prioritization and assignment.}

\item[Issue Management]{Exceptions trigger classes to auto-file bug reports against other classes instead of humans doing so. Classes, reading customer reviews, file feature requests against its libraries. When selecting a bug or feature to work on next, classes may prioritize those that are filed by classes with strong reputations or large follower/client counts. Classes tweet and send friend requests to developers who have the experience to fix or improve them. Developers follow the classes they developed and also the classes they are interested in. Developers no longer have to look for classes to work on or fix next --- the classes that are in need contact developers directly. In some cases, classes may fix themselves by consulting other relevant API drones and generating pull requests for human review. \textit{Intelligent} bug reports or bug drones directly contact the relevant developers reducing or completely eliminating bug \textit{tossing}. A bug drone collaborates with other code drones such as requirement and test drones choreographing efficient and (semi)autonomous solutions to bug \textit{triaging} and \textit{prioritization}.}

\item[Instant Reuse]{
With the classes from different teams in the company being \emph{aware} of each other's existence (because they are proactive and share the same company-wide agent cloud), reuse opportunities are discovered in near-real-time and relayed to the relevant stakeholders. Classes are aware of their purpose and may collaborate with each other and consult the Internet for authoring/updating comments and documentation.}

\item[Immersive IDEs]{Current IDEs such as Visual Studio and Eclipse will evolve from smart editors to virtual conference rooms in the social network. Developers and other code drones will meet in such conference rooms and collaborate to discuss and compose new software through touch screens \cite{muller:chase-icse13} and gestures. Editing code by hand will become a thing of past. The classes, harnessing the power of their agent cloud, will be able to converse and work with the developers using voice-based, VR-based~\cite{elliott:icse-nier15}, or CodeCity-like~\cite{codecity} interfaces during the otherwise mundane sessions such as debugging and refactoring. The ``crowd'' in crowd-sourced software engineering \cite{parnin:icpc13} will include humans and humanized artifacts (code drones).}

\item[Social Software Estimation]{Some key \textit{knowledge}/\textit{memory} embedded in a \texttt{Code Drones}-enabled artifact is the time needed for it to be created, as well as the originally estimated time. A class will also be aware of its own quality attributes such as its level of security and its performance. This knowledge is carried and updated throughout the life of the software artifact. When the artifact requires an update (a self-update or otherwise), it can initiate an estimation process that follows established estimation principles such as the \textit{cone of uncertainty} or history-based estimation. The software artifact can also contact other code drones that may have gone through a similar upgrade and thus have more reference points for its own estimation. The software artifact could also compute the expected \textit{velocity} based on the knowledge of who will be doing the upgrade (self or a developer).}

\item[Social Patching]{With Boa/Google/Facebook-like search, classes continually search Stack Overflow, national vulnerability databases, problem discussions in consumer forums, hacker exchanges, and repositories on the Internet to fix themselves and alert other relevant classes/developers in their social network via a private message or phone call.}

\item[Verification and Testing]{Classes use standard automated test generation APIs to generate test cases. Then they work with test drones and invoke change impact analysis APIs after changes for fully automated regression testing. Classes collaborate with each other to statically generate execution paths that violate a specification or dynamically generate system-level test cases that expose a security vulnerability.}


\item[Traceability]{A group of existing related classes in a company's system may contact a product manager directly after these classes read the requirements gathered by the manager from a customer interview published on the company Intranet. These classes directly negotiate requirements with that customer, on behalf of the product manager, and maintain up-to-date traceability links between requirements, code, and tests.}



\end{description}


\section{Discussion}
\label{sec:discuss}

A primary research challenge for enabling \texttt{Code Drones} is to model software engineering problems as MAS problems, i.e., in terms of autonomous interacting code drones who must sometimes coordinate and other times compete. For example, software engineering problems such as verifying specifications and integration testing need to be formulated as \textit{cooperative games}, while others such as building a reputation system for developers and humanized artifacts may map to non-cooperative games. Will a class be willing to \textit{die} if something better comes along? With software engineering problems modeled as MAS problems, implementing \texttt{Code Drones} may benefit from paradigms such as \textit{Agent-Oriented Software Engineering}~\cite{wooldridge:lncs02, aop} and \textit{Interaction-Oriented Software Engineering}~\cite{chopra:arxiv12}. 

Modeling software engineering problems as MAS problems could be a natural way to introduce principles of \textit{Distributed Computing}~\cite{distcomp} for traditional software engineering approaches that are not known to scale such as static verification and integration testing. In distributed computing, a problem is divided into many tasks, each of which is solved by one or more computers, which communicate with each other by message passing. Mapping testing and verification tasks to \texttt{Code Drones} agents may lead to truly distributed solutions to these expensive approaches. For instance, each agent can be mapped to a \textit{MapReduce}~\cite{dean:osdi04} job to take advantage of a distributed cluster. Does \texttt{Code Drones} enable \textit{Distributed software engineering (DSE)}, a paradigm where agents must not only use distributed computing resources, but must also coordinate actions across many artifacts and stakeholders? 




Crowd-sourced approaches exist for variety of software engineering tasks, ranging from documentation \cite{parnin:icpc13} and design to coding \cite{topcoder}, debugging \cite{bugfinders}, and testing \cite{crowdsourcetesting}. With a social platform such as the extended GitHub or Codebook, \texttt{Code Drones} nicely complements crowd-sourced software engineering in that it brings together humans, intelligent software artifacts (code drones), and human/artifact organizations into a single platform and federates the crowd efforts and the artifact-choreographed initiatives. The ``crowd'' in crowd-sourced software engineering will have both humans and humanized artifacts (code drones). How can a system or a design space be created that takes into account different dimensions of centralized control and self-service? How are policies, oversight, control, trusted ``news sources'', security guidelines, and verification set and vetted? 

In the Social Internet of Things (SIoT) \cite{siot} paradigm, billions of embedded computing devices form a social network. However, such social networks exist for humans to provide a structure to the Internet of Things (IoT) and efficiently navigate and access the results of the social interdevice communication. In \texttt{Code Drones}, unlike SIoT, classes, like humans, constantly seek opportunities for improvement using the power of the agent cloud and then publish such improvements to its social network. In doing so, classes choreograph efficient solutions to many software development and maintenance challenges. The classes view the social network of \texttt{Code Drones} as \textit{Internet of Code} (IoC) and humans for information exchange. In the not so distant future, it will not be surprising to see the seamless merging of the present day Internet, SIoT, and the social platform of \texttt{Code Drones}. What are the challenges and opportunities when the code that runs a sensor also becomes a part of IoT along with the sensor? 

What does the \texttt{Code Drones} paradigm mean to AI? First, as \texttt{Code Drones} agents themselves are artificially intelligent, even simple \texttt{Code Drones}-enabled programs such as text editors and calculators will be artificially intelligent. Second, consider the following. IBM Watson~\cite{watson} is artificially intelligent. Watson has some code behind it. Without \texttt{Code Drones}, over time, the Watson software learns based on runtime observations alone without any changes to its underlying code. With \texttt{Code Drones}, Watson code will have \texttt{Code Drones} agents, and these agents are artificially intelligent as well. This would mean that Watson will not only learn based on runtime observations, but also auto-evolve its code, gaining new capabilities over time. Will \texttt{Code Drones} mean more powerful AI systems?

\texttt{Code Drones} is not expected to compose new classes from scratch or entirely replace developer's creativity; for now. Developers will still have to handhold the classes initially and whenever a new functionality needs to be added. However, \texttt{Code Drones} will be handy in today's world where almost no one writes software from scratch. The \texttt{Code Drones} paradigm will liberate developers and maintainers from mundane activities such as refactoring for performance, finding opportunities for reuse in a company, testing, API migration, and following security best practices. Instead, developers will be able to focus their energies on creative activities such as creating more creative classes. In this paper, we discussed some initial directions and laid out the research challenges for realizing the vision of a \texttt{Code Drones}-enabled world. 

\bibliographystyle{abbrv}
\bibliography{sigproc}

\begin{thebibliography}{10}

\bibitem{aop}
Agent {O}riented {P}rogramming.
\newblock \url{https://en.wikipedia.org/wiki/Agent-oriented_programming}.

\bibitem{autocomp}
Autonomic {C}omputing.
\newblock \url{https://en.wikipedia.org/wiki/Autonomic_computing}.

\bibitem{boa}
Boa.
\newblock \url{boa.cs.iastate.edu}.

\bibitem{bugfinders}
Bugfinders.
\newblock \url{bugfinders.com}.

\bibitem{cmu_mas}
{CMU} intelligent software agents.
\newblock \url{cs.cmu.edu/~softagents/multi.html}.

\bibitem{codebook}
Codebook.
\newblock \url{research.microsoft.com/en-us/projects/codebook}.

\bibitem{codecity}
{CodeCity}.
\newblock \url{www.inf.usi.ch/phd/wettel/codecity.html}.

\bibitem{crowdsourcetesting}
crowdsourcetesting.
\newblock \url{crowdsourcetesting.com}.

\bibitem{devassist}
Developer {A}ssistant.
\newblock
  \url{visualstudiogallery.msdn.microsoft.com/a1166718-a2d9-4a48-a5fd-504ff4ad1b65}.

\bibitem{distcomp}
Distributed {C}omputing.
\newblock \url{https://en.wikipedia.org/wiki/Distributed_computing}.

\bibitem{watson}
{IBM} {W}atson.
\newblock \url{ibm.com/smarterplanet/us/en/ibmwatson}.

\bibitem{mas}
Multi-{A}gent {S}ystems.
\newblock \url{en.wikipedia.org/wiki/Multi-Agent_System}.

\bibitem{siot}
Social {I}nternet of {T}hings.
\newblock \url{social-iot.org}.

\bibitem{sna}
Social network analysis.
\newblock \url{en.wikipedia.org/wiki/Social_network_analysis}.

\bibitem{topcoder}
Topcoder.
\newblock \url{topcoder.com}.

\bibitem{ball:cav07}
T.~Ball and S.~Rajamani.
\newblock The {SLAM} toolkit.
\newblock In {\em CAV}, 2007.

\bibitem{becker:13}
K.~Becker.
\newblock Using artificial intelligence to write self-modifying/improving
  programs.
\newblock \url{http://www.primaryobjects.com/CMS/Article149}.

\bibitem{chaos-monkey}
C.~Bennett and A.~Tseitlin.
\newblock Chaos monkey released into the wild.
\newblock
  \url{techblog.netflix.com/2012/07/chaos-monkey-released-into-wild.html}, July
  2012.

\bibitem{bounimova:icse13}
E.~Bounimova, P.~Godefroid, and D.~Molnar.
\newblock Billions and billions of constraints: {W}hitebox fuzz testing in
  production.
\newblock In {\em ICSE}, 2013.

\bibitem{chopra:arxiv12}
A.~Chopra and M.~Singh.
\newblock {I}nteraction-{O}riented {S}oftware {E}ngineering.
\newblock {\em arxiv.org/abs/1211.4123}, 2012.

\bibitem{dean:osdi04}
J.~Dean and S.~Ghemawat.
\newblock Map{R}educe: {S}implified data processing on large clusters.
\newblock In {\em OSDI}, 2004.

\bibitem{elliott:icse-nier15}
A.~Elliott, B.~Peiris, and C.~Parnin.
\newblock Virtual reality in software engineering: Affordances, applications,
  and challenges.
\newblock In {\em ICSE NIER}, 2015.

\bibitem{jones:ase05}
J.~A. Jones and M.~J. Harrold.
\newblock Empirical evaluation of the tarantula automatic fault-localization
  technique.
\newblock In {\em ASE}, 2005.

\bibitem{kim:icse13}
D.~Kim, J.~Nam, J.~Song, and S.~Kim.
\newblock Automatic patch generation learned from human-written patches.
\newblock In {\em ICSE}, 2013.

\bibitem{mcburney:icpc14}
P.~McBurney and C.~McMillan.
\newblock Automatic document generation via source code summarization of method
  context.
\newblock In {\em ICPC}, 2014.

\bibitem{muller:chase-icse13}
S.~Muller, M.~Wursch, T.~Fritz, and H.~C. Gall.
\newblock An approach for collaborative code reviews using multi-touch
  technology.
\newblock In {\em CHASE at ICSE}, 2013.

\bibitem{parnin:icpc13}
C.~Parnin, C.~Treude, L.~Grammel, and M.~A. Storey.
\newblock Crowd documentation: {E}xploring the coverage and the dynamics of
  {API} discussion on {Stack Overflow}.
\newblock In {\em ICPC}, 2013.

\bibitem{wooldridge:lncs02}
M.~Wooldridge and P.~Ciancarini.
\newblock Agent-{O}riented {S}oftware {E}ngineering: {T}he state of the art.
\newblock In {\em LNCS Vol. 1957}, 2002.

\end{thebibliography}
\balancecolumns
\end{document}